*Preprint*
*Encyclopedia of Life Support Systems (EOLSS), UNESCO (2009)*

# PHASE BEHAVIOR IN PETROLEUM FLUIDS

(*A Detailed Descriptive and Illustrative Account*)

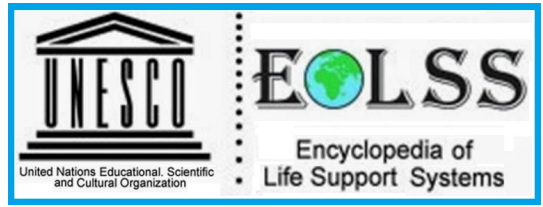

(2009)

G.Ali Mansoori

*Departments of BioEngineering, Chemical Engineering & Physics, University of Illinois at Chicago, Chicago, IL 60607-7052* USA   Tel.: +1-312-996-5592; email: mansoori@uic.edu

Contents

1. Introduction
    1.1. Naturally Occurring Petroleum Fluids
2. Components of Petroleum Fluids
    2.2. Impurities in Petroleum Fluids
    2.3. Heavy Fractions in Petroleum Fluids
        2.3.1. Petroleum Wax
        2.3.2. Diamondoids
        2.3.3. Asphaltenes
        2.3.4. Petroleum Resins
3. Phase Behaviors in Petroleum Fluids
    3.1. Temperature Effect on Petroleum Fluids Phase Separation
    3.2. Pressure Effect on Petroleum Fluids Phase Separation
    3.3. Theory of Phase-Transitions
    3.4. Phase-Transition Points
4. Discussion
Clossary
Bibliography

**Summary**

This chapter presents a descriptive and illustrative account of phase behavior in the seven naturally occurring petroleum fluids and ties all the known eleven phase-transition concepts in a unified narrative. The figures and tables contained in this report are designed so that they could effectively support the discussion about molecular make-up of petroleum fluids, P- and T-effects on phase behavior and phase transition points.

Seven naturally occurring hydrocarbon fluids are known as petroleum fluids. They include, in the order of their fluidity, natural gas, gas-condensate (NGL), light crude, intermediate crude, heavy oil, tar sand and oil shale. In this report we present a generalized description of the various phase transitions, which may occur in petroleum fluids with emphasis on heavy organics deposition..

At first the nature of every petroleum fluid is presented. Their constituents including their so-called impurities are identified and categorized. Heavy fractions in petroleum fluids are discussed and their main families of constituents are presented including petroleum wax, diamondoids, asphaltenes and petroleum resins. Then the generalized petroleum fluids phase behavior is discussed in light of the known theory of phase transitions. The effects of variations of composition, temperature and pressure on the phase behavior of petroleum fluids are introduced. Finally eleven distinct phase-transition points of petroleum fluids are presented and their relation with state variables and constituents of petroleum fluids are identified. This report is to generalize and relate phase behaviors of all the seven naturally occurring petroleum fluids into a unified perspective. This work is the basis to develop a comprehensive computational model for phase behavior prediction of all the petroleum fluids, which is of major interest in the petroleum industry.

## 1. Introduction

The petroleum and natural gas literature is quite rich in data, industrial correlations and molecular-based prediction methods of the liquid-vapor phase behavior of petroleum fluids (see [1]-[17] ). For this reason little effort is made to discuss further liquid-vapor equilibrium in this report. Our emphasis is to introduce a generalized perspective about all the variety of phase transitions, which may occur in the seven category of naturally occurring petroleum fluids. This includes vapor separation, solid crystalline deposition, colloidal and micellar solutions formation as well as aggregation, flocculation and non-crystalline solid formation and separation. We are in the process of developing a comprehensive phase behavior prediction package for petroleum fluids, which can be applied to any of the seven categories of such fluids from underground natural reservoirs. Having a thorough understanding of all the possible phase transitions in the seven categories of petroleum fluids will allow us to formulate the necessary computational package for their phase behavior prediction regardless of the kind, source, nature of components and thermodynamic conditions.

### 1.1. Naturally Occurring Petroleum Fluids

There exist seven well-known petroleum fluids in nature. In the order of their fluidity, they are natural gas, gas-condensate (also known as NGL standing for natural-gas liquid), light crude, intermediate crude, heavy oil, tar sand and oil shale as shown symbolically in Figure 1. These are all naturally occurring complex mixtures made up of hydrocarbons and other organic and inorganic compounds with variety of molecular structures and sizes.

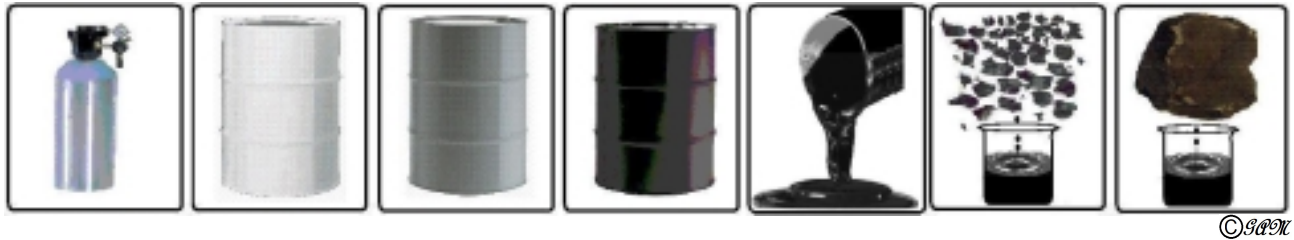

**Figure 1** – *The normal conditions of the seven naturally occurring petroleum fluids, in the order of fluidity from left to right, natural gas, gas-condensate (NGL), light crude oil, intermediate crude oil, heavy oil, tar sand and oil shale.*

The hydrocarbons and most other organic compounds present in all these seven naturally occurring petroleum fluids are generally polydispersed having a range of size, shape and molecular weight distributions (see for example [18,]). As a result continuous functions may be used for their mole fraction distributions (see for example [19-24]). In Figure 2 we report the approximate relative composition of various petroleum fluids with respect to the number of carbon atoms of their organic contents.

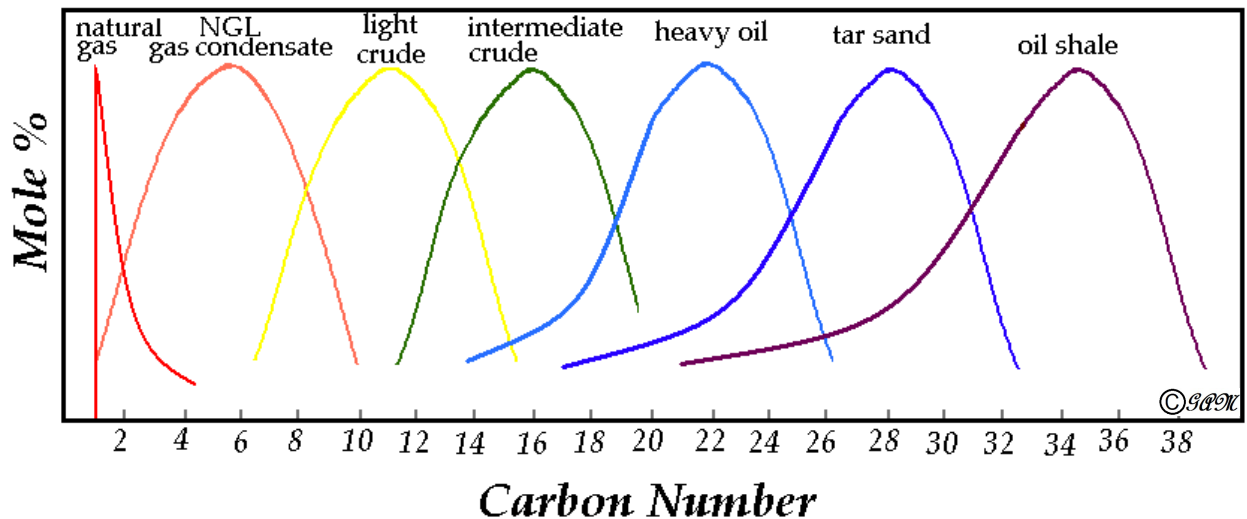

**Figure 2** – *Various categories of natural gas and liquid hydrocarbon reserves and their approximate hydrocarbon molecular weight distributions according to their carbon numbers*

We think this is a useful way to classify petroleum fluids according to their composition. This novel classification will allow us to look at all the petroleum fluids from a unified basis which is their hydrocarbon contents and their relative distribution. According to this figure, the molecular weight of hydrocarbons present in petroleum fluids increases from left to right in the list of the seven. It should be also mentioned that heavy oil, tar sand and oil shale contain appreciable amounts of resins and asphaltenes, while the percentages of these compounds are generally low in the lighter petroleum fluids.

Petroleum fluids vary in color, odor, and physicochemical properties. Generally many compounds are known to be present in petroleum fluids. The number of carbon atoms in the hydrocarbons and other organic compounds present in petroleum fluids can vary from one (as in methane) to over a hundred (as in asphaltenes and heavy paraffins) [25,26].

During their production, transportation and processing petroleum fluids may go through a number of phase changes, which include evaporation (separation of gases from liquid streams), retrograde condensation (separation of liquids from gas streams) and solids formation and deposition (separation of crystalline solids, colloids and aggregates from liquid or vapor streams). The latter is mostly due to precipitation and / or deposition of diamondoids, wax and asphaltenes. In view of the complexity of petroleum fluids, study and understanding of their phase behavior is still a challenging and an industrially important task. Such an understanding will help us to design a more economical route for the related production, transportation and refining projects. The complexity of the phase behavior in petroleum fluids is due to the existence of the variety and polydispersity of hydrocarbons and other organic molecules in them. In this report, we present the various phase-transitions, which may occur in petroleum fluids, and we introduce a unified perspective of their phase behaviors. This may allow us to develop a comprehensive theoretical model for phase behavior prediction of all the petroleum fluids which is of major concern in the petroleum industry.

## 2. Components of Petroleum Fluids

The predominant hydrocarbons present in lighter petroleum fluids (natural gas, gas-condensate, light-crude and intermediate-crude oil systems) are alkanes (also known as paraffin hydrocarbons) as shown in Figure 3. As we look at the physical properties of petroleum fluids, going from natural gas to oil shale, their alkanes content decrease while their viscosity and density increase. Other hydrocarbons present in petroleum fluids are aromatic hydrocarbons (see Figure 4) at somehow lower concentrations than paraffins depending on the underground natural reservoir source. Other families of hydrocarbons are also present in petroleum fluids but their concentrations are generally rather low.

Quite frequently, petroleum fluids contain various amounts of other organic and inorganic compounds, which are usually termed as impurities from the petroleum industry point of view.

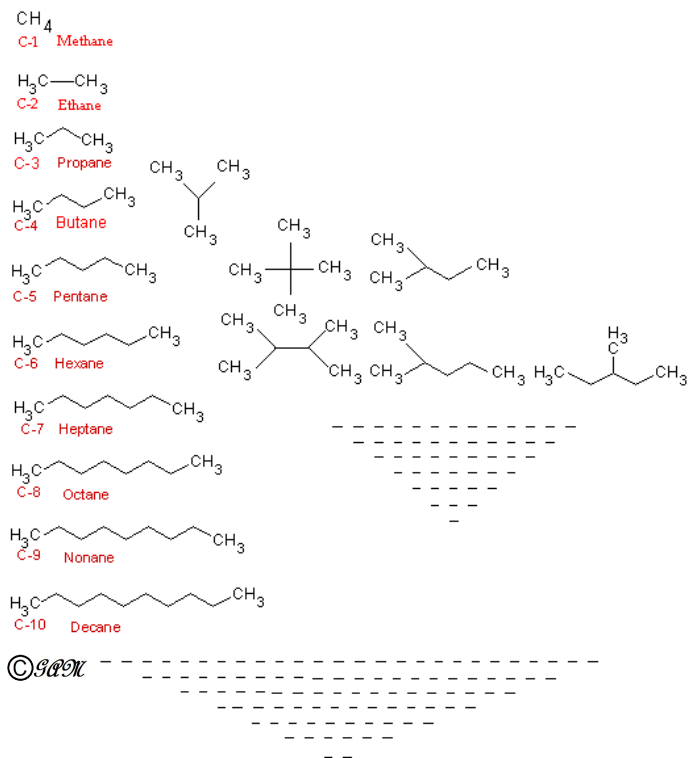

**Figure 3** - *Lighter paraffin hydrocarbons present in petroleum fluids*

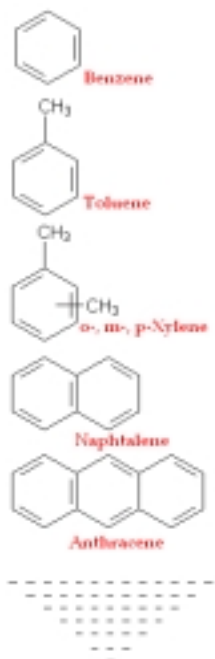

**Figure 4** - *Lighter aromatic hydrocarbons present in petroleum fluids*

## 2.2. Impurities in Petroleum Fluids

All the non-hydrocarbon organic and inorganic components of petroleum fluids are considered impurities from the petroleum industry perspective. Molecular structure of organic impurities is composed of mostly such elements as carbon, hydrogen, metals, nitrogen, oxygen and sulfur. Molecules of inorganic impurities are generally composed of carbon, metals, oxygen and sulfur. Petroleum fluid impurities can be categorized into five groups of compounds:

(i). The major low molecular weight impurities, which include carbon dioxide ($CO_2$), hydrogen sulfide ($H_2S$), metal oxides ($Al_2O_3$, $Fe_2O_3$, $SiO_2$, etc.), nitrogen ($N_2$), oxygen ($O_2$), salts ($NaCl$, $CaCO_3$, etc.), sulfur ($S$) and water ($H_2O$).

(ii). High molecular weight impurities, which could be present in the petroleum fluids heavy fractions. They include asphaltenes, asphaltogenic acids, diamondoids and derivatives, mercaptans, metal carbenes, organometallics, petroleum resins and wax. In Table I the general closed chemical formulae of high molecular weight impurities of petroleum fluids are presented.

(iii). There may be also other organic and inorganic compounds present in petroleum fluids, which depend on their underground natural reservoir source. The concentration of such other organic matter is generally rather low and they have little or no effect on petroleum fluids phase behavior.

**Table 1**: *General closed chemical formulae of high molecular weight impurities of petroleum fluids. In this table C is carbon, H is hydrogen, N is nitrogen, O is oxygen, S is sulfur and X is metals and R represents a hydrocarbon segment.*

| High Mw Impurity | General Closed Formula |
|---|---|
| Asphaltenes | $C_mH_nN_iO_jS_k$ |
| Asphaltogenic Acids | $C_mH_nN_iO_jS_k$-COOH |
| Diamondoids | $C_{4n+6}H_{4n+12}$ |
| Mercaptans | $C_mH_nN_iO_jS_kX$ |
| Metal Carbenes | $R_1HC=CHR_2X$ |
| Organometallics | Ex.: $R_1$-X-$R_2$ |
| Petroleum Resins | $C_mH_nN_iO_jS_k$    $i,j,k=0$ or $1.0$ |
| Wax | $C_mH_n$    $18 \leq m \leq 60$   $n \leq 2m+2$ |

(iv). Another category of impurities which may be present are compounds which have been added to petroleum fluids during their production, transportations and storage stages for various reasons. These include but not limited to acids, alcohols, aromatic hydrocarbons, detergents and polymers.

(v). Petroleum fluids often contain compounds, which are a result of physical association of hydrocarbons and the above-mentioned impurities and contaminants. They include clatherates, colloids, crystalline solids, flocs and slugs.

As it was mentioned above the literature on petroleum and natural gas is rich of numerous data banks, correlations, and nomograms for the analysis and treatment of petroleum fluids and their liquid-vapor phase behavior. The presentation given in this report is an attempt to relate properties and phase behavior of petroleum fluid systems to the molecular behavior of the compounds present in them. This would allow us to make a unified and a science-based generalization of the behavior of petroleum fluids.

One of the important fractions of petroleum fluids, which is the least understood is their heavy fraction. Because of the complexity of the heavy fractions, we discuss them in more detail in the following separate section.

**2.3. Heavy Fractions in Petroleum Fluids:**

To perform an accurate prediction of the phase behavior of petroleum fluids we need to know the nature and composition of the molecules, which make up the petroleum heavy fractions [27]. There is a wealth of information and database available for the light and intermediate components of petroleum fluids [1-17]. The investigation of the accurate chemical constitution of petroleum heavy fractions is hindered by their complex nature. Almost all the molecules comprising the heavy fractions of petroleum fluids are polydispersed [18,27,28,29]. The predominant part of the high molecular weight impurities (known as heavy fractions or petroleum residuum) are asphaltenes, asphaltogenic acids/compounds, diamondoids and derivative, heavy aromatic hydrocarbons, mercaptans, metal carbenes / organometallic, petroleum resins and wax. In Table 1 we report the general closed chemical formulae of these compounds.

It is generally understood that heavy fractions have little or no effect on the liquid-vapor phase behavior of the majority of petroleum fluids. Their main contribution is in the solids separation from petroleum fluids, due to changes in the composition, temperature and pressure. The main components of the heavy fraction, which participate in the solid phase formation, include asphaltenes, diamondoids, petroleum resins and wax. In what follows we present a more detailed description of their molecular characteristics and their physical properties.

**2.3.1. Petroleum Wax:** Petroleum wax is a class of mineral wax that is naturally occurring in the heavy fractions of petroleum fluids [30-37]. They vary compositionally over a wide range of molecular weights up to hydrocarbon chain lengths of approximately $C_{60}$ [34]. Petroleum wax is made up mostly of saturated paraffin hydrocarbons with their number of carbon atoms in between 18-36 (see Figure 5).

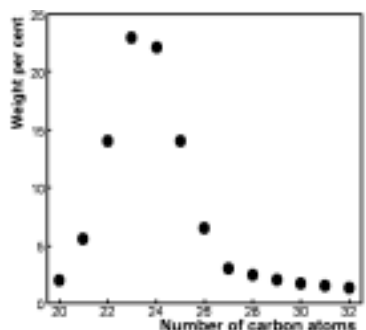

**Figure 5 -** *The distribution of n-alkanes in Suntech P116 paraffin wax as a function of the number of carbon-atoms (Taken from [32])*

Wax may also contain small amounts of naphthenic hydrocarbons with their number of carbon atoms in the range of 30-60. Wax usually exists in intermediate crudes, heavy oils, tar sands and oil shales. Petroleum wax is typically in solid state at room temperature and it is separated from relatively high boiling petroleum fractions during the refining processes. Petroleum wax, like most other components of heavy end of petroleum fluids, is polydispersed. As an example, we report in Figure 5 the distribution of n-alkanes as a function of the number of carbon-atom compounds in a paraffin wax sample.

**2.3.2. Diamondoids:** Diamondoid molecules are cage-like ultrastable saturated hydrocarbons [38-40]. These molecules are ringed compounds, which have a diamond-like structure consisting of a number of six-member carbon rings fused together (see Figure 6).

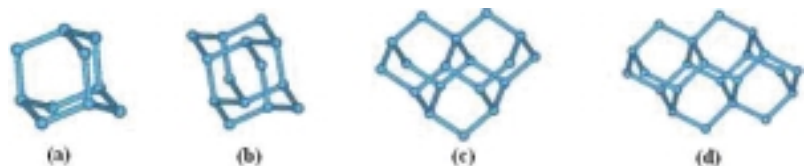

**Figure 6 -** *Molecular structures of Diamondoids: (a): Adamantane (b): Diamantane (c): Triamantane (d): Anti- isomer of Tetramantane.*

In other words, they consist of repeating units of ten carbon atoms forming a tetracyclic cage system [38]. They are called "diamondoid" because their carbon-carbon framework constitutes the fundamental repeating unit in the diamond lattice structure. Bragg and Bragg using X-ray diffraction analysis first identified them in 1913. Diamondoids are constituents of petroleum, gas-condensate and natural gas reservoirs. Adamantane, the smallest diamondoid molecule was originally discovered and isolated from petroleum fractions of the Hodonin oilfields in Czechoslovakia in 1933. Naturally occurring adamantane is generally accompanied by small amounts of alkylated adamantane: 2-methyl-; 1-ethyl-; and probably 1-methyl-; 1,3-dimethyl-adamantane; and others. Diamantane, triamantane and their alkyl-substituted compounds, just as adamantane, are also present in certain petroleum crude oils. Their concentrations in crude oils are generally lower than that of adamantane and its alkyl-substituted compounds.

**2.3.3. Asphaltenes:** Asphaltenes are high molecular weight polycyclic organic compounds with nitrogen, oxygen and sulfur in their structure in addition to carbon and hydrogen [30,31,33,41-50]. Asphaltene present in petroleum fluids is defined as the fraction of petroleum fluid (or other carbonaceous sources such as coal), which is soluble in benzene and deposits, by addition of a low-boiling paraffin solvent. Asphaltogenic acids and related compounds may be present in petroleum usually in insignificant quantities. In Figures 7 we report four different asphaltene structures separated from different natural petroleum fluids.

Asphaltene is not crystallized upon deposition from petroleum fluids and as a result, its phase-transition from liquid to solid does not follow the same route as paraffin wax. Asphaltene cannot be easily separated into individual purified components or narrow fractions. Thus, the ultimate analysis is not very significant, particularly taking into consideration that the resins are strongly adsorbed by asphaltenes and could not easily be quantitatively separated from them.

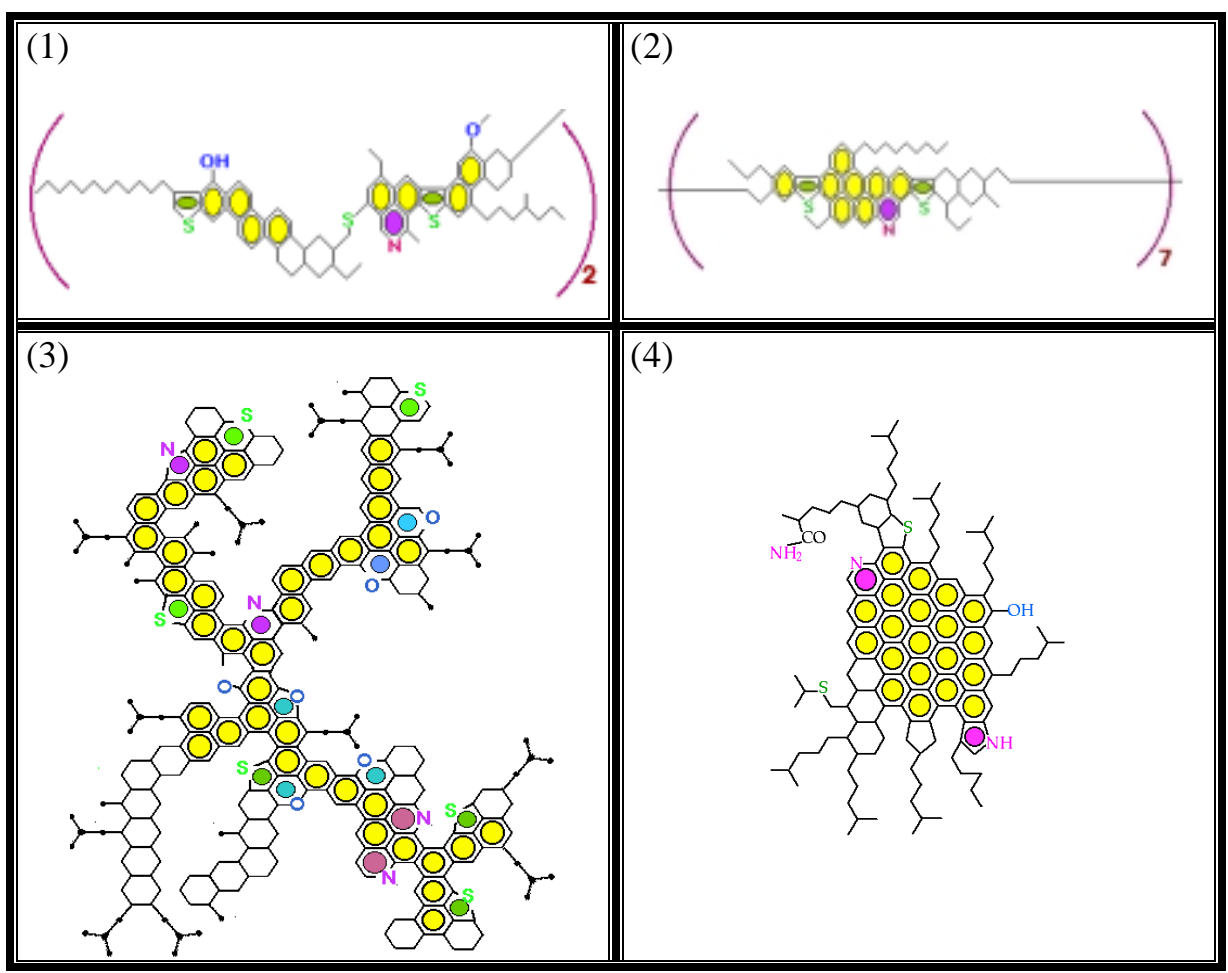

**Figure 7** - *Average molecular structure models of: (1) asphaltene fraction of Athabasca tar–sand (Canada) [48]; (2) asphaltene fraction of Athabasca heavy oil (Canada) [48]; (3) asphaltene*

*proposed for Maya crude (Mexico) [49]; (4) asphaltene Proposed for 510C Crude Residue (Venezuela) [50]. In these figures ⬢ represents benzene ring an ⬡ represents cyclohexane ring.*

Asphaltene particles can assume various forms when mixed with other molecules depending on the relative sizes and polarities of the particles present. Asphaltenes are lyophilic with respect to aromatics, in which they form highly scattered reverse-micelle solutions [51-53]. The color of dissolved asphaltenes in aromatics (like toluene) varies from yellow at low concentrations to deep red at higher concentrations when they form micelles. On the other hand, asphaltenes are lyophobic with respect to low molecular weight paraffins, in which they form scattered flocs, which may form steric colloids due to the presence of resins in the solution [33,54] (see Figure 8). On heating, asphaltenes are not melted, but decompose, forming carbon and volatile products above 300-400 ºC.

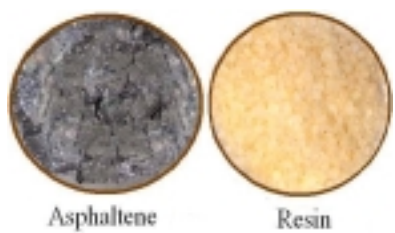

**Figure 8.** *The asphaltene sample shown here is actually made of asphaltene deposits consisting of asphaltene steric colloid flocs and encapsulated in them the other heavy organics. The resin sample shown here is a purified petroleum resin.*

We must recognize the fact that petroleum asphaltene may form micellar particles in aromatic and/or polar solvents such as toluene and methyl-naphthalene [51-53, 55]. On the other hand, asphaltene may form steric-colloidal particles in the presence of excess amounts of light alkanes (paraffin hydrocarbons) and resins [56-60]. Small-size asphaltene particles may be dissolved in a petroleum fluid, whereas relatively large asphaltene particles may be separated from petroleum fluids, due to high paraffin content of the oil, forming random aggregates. These aggregates may be micellized by the addition of aromatics or can be colloidally dispersed in the presence of appropriate resins. The color of petroleum fluids and residues is affected due to presence of petroleum resins and asphaltenes. The black color of some crude oils and residues is related to the presence of asphaltenes, which are not well peptized.

There are efforts underway to characterize asphaltenes in terms of their chemical structure, elemental analysis and carbonaceous sources. A number of investigators have postulated model structures for asphaltenes based on physical and chemical methods. Physical methods include IR, NMR, ESR, mass spectrometry, x-ray, ultra centrifugation, electron microscopy, small angle neutron scattering, small angle x-ray scattering, quasi elastic light scattering spectroscopy, VPO, GPC, HPLC, etc. Chemical methods involve oxidation, hydrogenation, etc. [47].

The asphaltene structures reported in Figure 7 are based on a number of physical and chemical methods mentioned above. All these structures contain carbon, hydrogen, oxygen, nitrogen, sulfur, as well as polar and nonpolar groups. As it is shown in Figure 7, asphaltene molecules have a high degree of polynuclear aromaticity. The length of paraffinic straight-chain of Athabasca asphaltenes is slightly longer than that of heavy Venezuelan petroleum asphaltene. Apparently, the degree of condensation of the aromatic ring system of Athabasca tar–sand asphaltene is considerably lower than that of Athabasca heavy oil asphaltene.

There have been considerable efforts to characterize the asphaltenes in terms of a combination of elemental analysis as well as by the carbonaceous sources. Asphaltenes in petroleum fluids are highly polydispersed with large molecular weight distributions. The average molecular weight of asphaltenes is very high ranging from approximately 1,000 to 2,000,000. Reported molecular weight of asphaltene varies considerably depending upon the method of measurements. A major concern in reporting molecular weights is the association of asphaltenes, which can exist at the conditions of the method of measurements. In Figure 9 we report an example of the molecular weight distribution of an asphaltene deposited from a crude oil by diluting the oil using three different paraffin hydrocarbons [18].

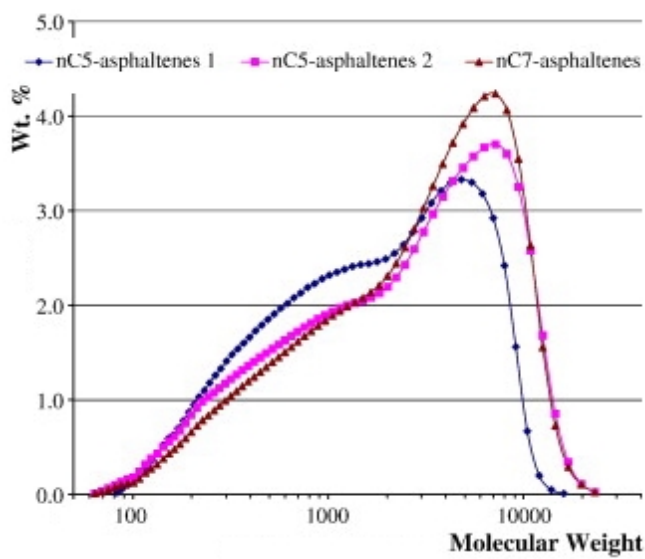

**Figure 9 -** *The molecular weight distribution of an asphaltene deposited from a petroleum fluid by using three different normal paraffin hydrocarbons and measured by gel permeation chromatography [18].*

**2.3.4. Petroleum Resins:** While the molecular weight of resins is much lower than those of asphaltenes, there is a close relationship between the molecular structures of asphaltenes and resins [33,41,61]. Figure 10 shows two representative structures for resin molecules belonging to the Athabasca tar sand and crude oil [48]. By comparing these figures with the Athabasca asphaltenes, shown in Figures 7.1 and 7.2, the similarity of structures and the big difference in their molecular weights become evident. There is a hypothesis on oxidation of resins as the precursor for asphaltene formation in nature.

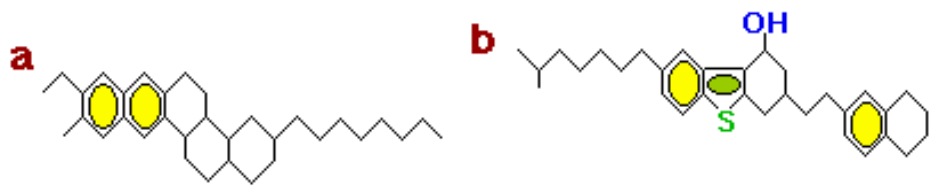

Figure 10 - *Average molecular structure models of: (a) resin fraction of Athabasca tar–sand and; (b) resin fraction of Athabasca petroleum [48].*

Structure of resins, as well as asphaltenes, can vary from one source to another. The physical and physicochemical properties of resins are different from those of asphaltenes. Petroleum resins are the main factor by which the asphaltenes remain dispersed in the oil medium when the oil is low in aromaticity. When resins are present in sufficient concentration in petroleum fluids their peptizing effect will prevent asphaltene self-association and it will delay asphaltene deposition.

Petroleum resin is completely miscible with light fractions of petroleum fluids, including light gasoline and petroleum ether. This is in contrast with asphaltene, which is insoluble in these light fractions, and as a result, it will precipitate out of the oil. Like other heavy organics present in the heavy fractions, petroleum resins generally posses a distribution of molecular weight and polarities. In Figure 11 we report an example of the molecular weight distribution of resins corresponding to the deposited asphaltene from the same petroleum fluid as in Figure 9 and measured by gel permeation chromatography [18]. There are efforts underway to characterize petroleum resins in terms of their chemical structure, elemental analysis and carbonaceous sources as for asphaltene.

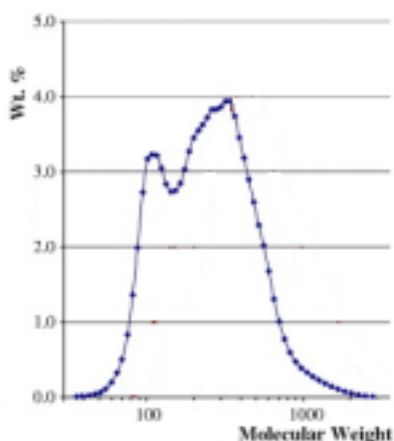

**Figure 11 -** *The molecular weight distribution of resins corresponding to the deposited asphaltene from the same petroleum fluid as in Figure 9 and measured by gel permeation chromatography [18].*

In the following section we present a detailed discussion of various phase-transitions, which may occur in petroleum fluids with their related, driving force variables, components involved and experimental indicator methods to observe and quantify them.

## 3. Phase Behaviors in Petroleum Fluids

Phase transitions or phase changes in petroleum fluids, which are usually referred to as "petroleum fluid phase behavior" in the petroleum and natural gas industries, are the transformation of a petroleum fluid from one phase to another. Due to their complexity and multi-family-component nature, a petroleum fluid could go through various phase-transitions, which are discussed below. Some of those phase-transitions are considered reversible and some are irreversible [62]. When reversible it is in the category of phase equilibrium. The general distinguishing characteristic of a phase transition is an abrupt change in one or more physical properties of the petroleum fluid.

All the seven petroleum fluids, which include natural gas, gas-condensate, light crude, intermediate crude, heavy oil, tar sand and oil shale, are in the category of complex mixtures. We define a complex mixture as one in which various families of compounds with diverse molecular properties are present. In petroleum fluids, there exist various families of hydrocarbons (paraffins, aromatics, naphthenes, diamondoids, etc.), various heavy non-hydrocarbon organic compounds (asphaltenes, resins, etc.) and other impurities (such as carbon dioxide, hydrogen sulfide, oxides, water, salts, etc.). Knowledge about the phase behavior of such complex mixtures is of interest in all sectors of the petroleum and natural gas industry including in the transportation, processing and refining industries.

Heavy organics deposition during petroleum fluids production, transportation and processing is a serious problem in many areas throughout the world. The economic implications of this problem are tremendous considering the fact that a problem workover cost each time could get as high as several million dollars. For example, in a heavy crude production field, formation of asphaltenic sludge after shutting in a well temporarily and/or after stimulation treatment by acid has resulted in partial or complete plugging of the well [63]. The downtime, cleaning and maintenance costs are a sizeable factor in the economics of producing a heavy crude field prone to organic deposition. Considering the trend of the oil industry towards the utilization of heavier asphaltenic crudes, heavy oil, tar sand and oil shale and the increased utilization of miscible flooding techniques for recovering and transportation of oil, the role of heavy organics deposition in the economic development of petroleum fluids production will be important and crucial.

At the production, transportation and refining conditions of petroleum fluid systems only hydrocarbons and the other organic compounds present in oil would participate in oil phase-transitions. Another word, inorganic compounds do not generally participate in such phase-transitions. In addition, in those same conditions it is understood that heavy organics will appear only in the solid and liquid phases and they will have the major role in solid formation during gas-solid and solid-liquid phase-transitions.

The major factors that govern deposition of heavy organics from petroleum fluids appear to be:

   i. Paraffins / wax crystallization because of lowering of temperature below the oil cloud point.
   ii. Asphaltene flocculation as initiated through variations in composition of petroleum fluid mixed with the injection (or blending) fluid, pressure and temperature.

With alterations in these parameters, the asphaltene flocculation and, as a result, the nature of heavy organics, which precipitate, will vary. In addition, it is a proven fact that the flocculation of asphaltene is generally followed by its deposition resulting with precipitates which are insoluble in petroleum fluids and containing other heavy organics and mineral deposits. We should also mention that diamondoids and some other heavy organics depositions due to various specific factors have also become a problem in recent years, some of which are under investigation [38].

The heavy organic compounds predominantly present in the solid phase separated from a petroleum fluid are paraffins, diamondoids, asphaltenes and resins. While other compounds and other heavy organics, including aromatics, may not participate in solid phase formation, they have a role in polarity of the oil medium and also may be encapsulated in the solid crystalline grid and / or aggregates.

**3.1. Temperature Effect on Petroleum Fluids Phase Separation:**

When we start with a petroleum fluid in liquid state, like the ones with a typical PT diagram as shown on Figure 12, by increasing its temperature, at constant pressure, we will observe the bubble point which is the start of evaporation of the volatile components of the petroleum. Upon further increase of petroleum fluid temperature we will observe the dew point after which all the volatile components of petroleum will go to the gas phase. By reversing this process the petroleum fluid will go back to its original liquid state. Liquid-vapor phase transition in petroleum fluids is, for all practical purposes, reversible.

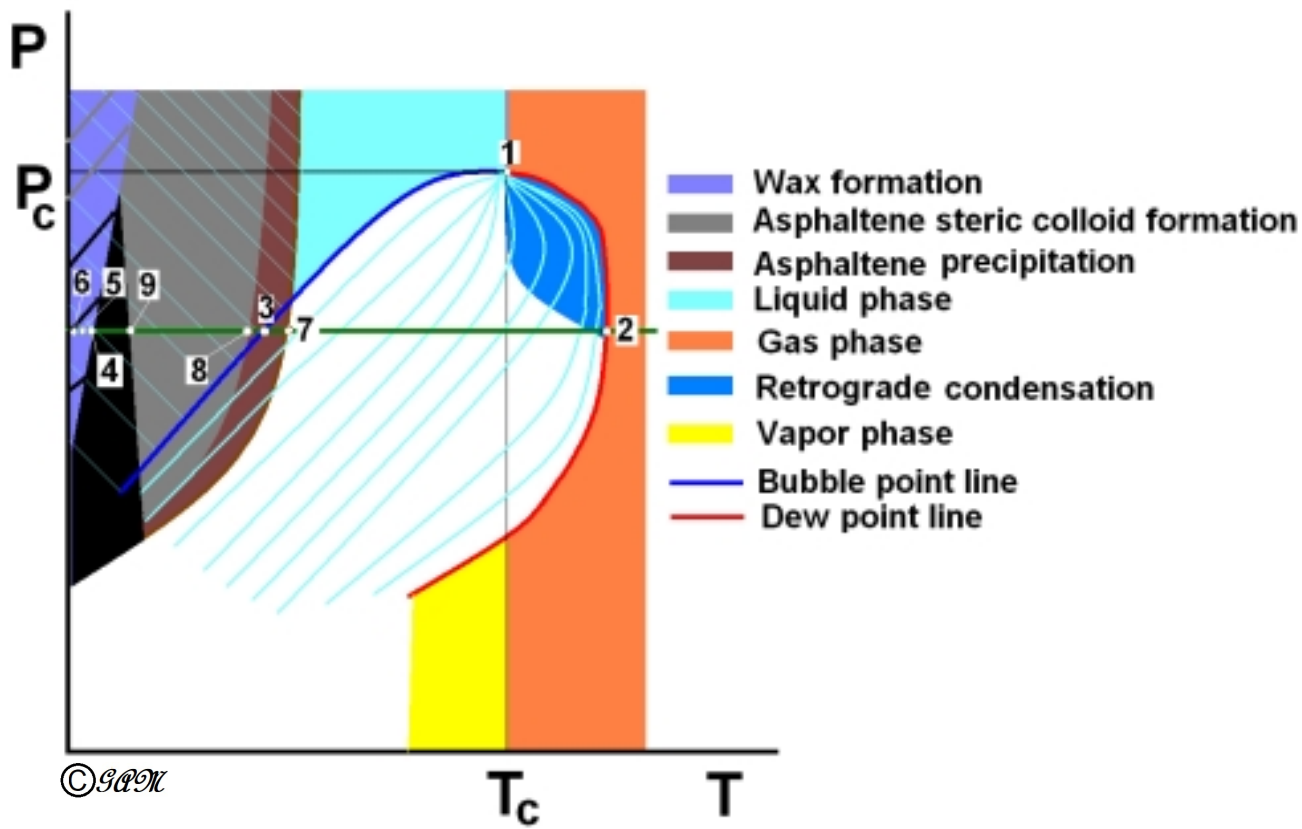

**Figure 12** - *A typical PT diagram of a petroleum fluid which may experience variety of phase transitions. This diagram as a whole may not apply to all the seven petroleum fluids, but it gives a general picture of how various phases and phase transitions may be distinguished from one another. In this figure points 1-9 correspond to the phase transition points 1-9 in Table 2.*

By an appreciable lowering of the temperature of a petroleum fluid originally in its liquid state, we could cause primarily; the precipitation and deposition of the wax contents of the petroleum fluid (see Figure 13). The wax present in petroleum fluids consists of paraffin hydrocarbons ($C_{18} – C_{36}$) known as paraffin wax and naphthenic hydrocarbons ($C_{30} – C_{60}$). Hydrocarbon components of wax can exist in various states of matter (gas, liquid or solid) depending on their temperature and pressure. When the wax freezes, it forms crystals. The crystals formed of paraffin wax are known as macrocrystalline wax (see Figures 14 and 15). Those formed from naphthenes are known as microcrystalline wax. Generally, wax would separate from petroleum fluids in solid form due to lowering of temperature. Mechanism of wax separation from a waxy crude oil is different from that of other petroleum fluids.

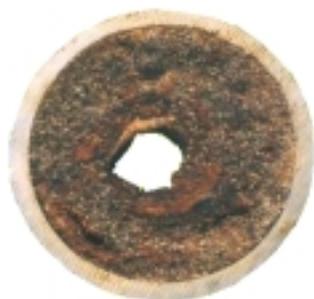

**Figure 13 -** *Pipeline petroleum transport plugging due to wax and other heavy organics depositions (Courtesy of Phillips Petroleum Company)*

Waxy crude is a petroleum fluid in which there exist only hydrocarbons including paraffins, aromatics and wax as its constituents. Solidification of heavy paraffins in wax and their crystallization will occur because of lowering the temperature of a petroleum fluid below its cloud point. Such fluid-solid transitions are generally reversible and the Gibbs equilibrium criteria apply to such transitions provided there are no other compounds like asphaltenes and resins present in the petroleum fluid. Presence of such other heavy organics could enhance or delay solidification of paraffins / wax depending on the form (molecular, aggregate, steric colloid. floc or micelle) of the asphaltene and resins and their molecular structures.

As a waxy crude flows through a cold pipe or conduit (with a wall temperature below the cloud point of the crude), crystals of wax may be formed on the wall (see Figures 14 and 15). Wax crystals could then grow in size until completely wax deposits cover the inner wall.

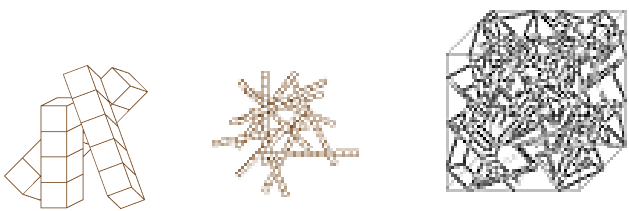

**Figure 14 -** *Macrocrystalline, Microcrystalline, and Crystal Deposit Network of Wax*

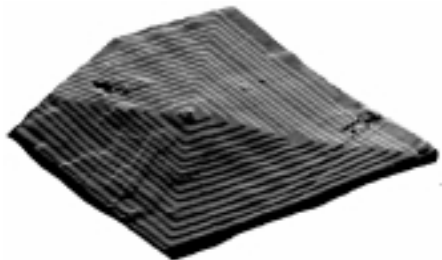

**Figure 15 -** *An atomic force microscope image of the spiral growth of paraffin crystal (measuring approximately 15 microns across). Inset shows orthorhombic arrangement (0.49nm x 0.84nm) of chain ends of one of the crystal terraces. (Courtesy of Professor M.J. Miles, Bristol University, GB).*

As the wax crystal thickness increases, pressure drop across the pipe needs to be increased to maintain a constant flow rate. As a result, the power requirement for the crude transport will increase. When the temperature of a waxy crude oil is lowered to its cloud point, first the heavier fractions of its wax content start to freeze out. Upon lowering of the temperature to the crude pour point almost all the fractions of its wax content will freeze out. Presently, a waxy crude is characterized by its cloud point and it static pour point, which are measured according to the ASTM Test Methods D-2500 and D-97, respectively.

As with regard to the other petroleum fluids their heavy organics deposition problem is not just crystallization of their wax content at lower temperatures, but the formation of deposits which do not disappear upon heating and will not be completely removed by mechanical means. Petroleum fluids, generally, in addition to wax, contain other heavy organics such as asphaltenes and resins. Asphaltenes do not crystallize upon cooling and, for the most part, do not have a definite freezing point. Depending on their nature, the other heavy organics will have different interactions with wax, which could either prevent wax crystal formation or enhance it.

### 3.2. Pressure Effect on Petroleum Fluids Phase Separation:

When we start with a petroleum fluid in liquid state by decreasing its pressure, at constant temperature, we will observe the bubble point, which is the start of evaporation of the volatile components of the petroleum. Upon further decrease of petroleum fluid pressure we will observe the dew point after which all the volatile components of petroleum will go to the vapor phase. By reversing this process the petroleum fluid will go back to its original liquid state. It must be pointed out that for some petroleum fluids, like the ones with a typical PT diagram as shown on Figure 12, if we change the pressure of the fluid at a temperature above the critical temperature we will also experience retrograde condensation [64].

By variations of pressure and / or composition of a petroleum fluid, we may cause the deposition of its polar heavy end content known as asphaltene. A petroleum fluid is a mixture of various polydispersed compounds. A mutual solubility balance among all such components makes a petroleum fluid stable. Changes in that balance can cause phase separations in a petroleum fluid and, as a result, formation of various other phases.

For example, dilution of petroleum fluids with low molecular weight alkanes could cause asphaltene precipitation, aggregation / flocculation and eventually deposition (see Figure 16) [65].

However, a paraffinic and asphaltenic petroleum fluid at low temperatures may not experience any solid formation due to the possible role that paraffin nanocrystals may play in peptization of asphaltene aggregates and formation of relatively stable steric colloids in the oil. For this kind of physical aggregation / flocculation, we have considered a polymerization chemical reaction mechanism along with the FRACTAL growth of aggregates [60,65,66] with an appropriate non-Euclidian dimensionality. In the absence of any peptizing agent, like petroleum resins, this aggregation will continue towards formation of a sticky non-crystalline solid phase. We have used the kinetic theories of aggregation and flocculation to predict solid formation of this nature.

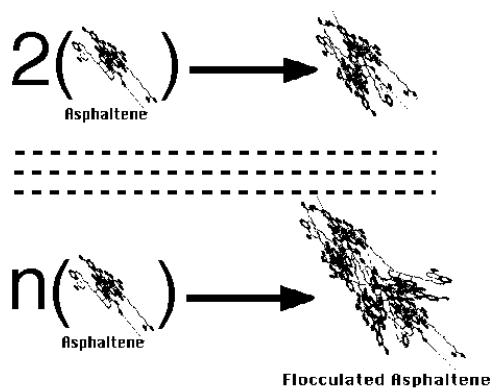

**Figure 16 -** *Aggregation / flocculation of asphaltene molecule [53]*

Asphaltene flocs (random aggregates) can form steric–colloids in the presence of petroleum resins [60,65,66], as shown in Figure 17.

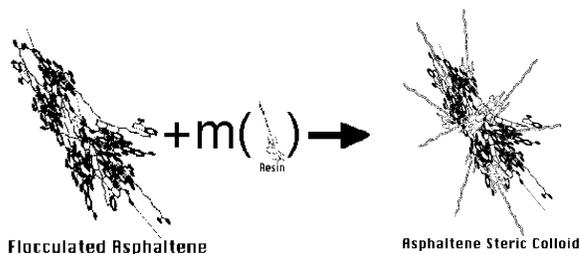

**Figure 17 -** *Steric colloid formation of asphaltene aggregate plus resins [53]*

The concept that asphaltene molecules could be present as a colloidal system is credited to Nellensteyn [67]. He proposed that asphaltenic compounds were made of flocs or aggregates of asphaltene molecules protected by adsorbed resin and hydrocarbon materials, all dispersed in a hydrocarbon medium. He also found that the peptizing or precipitating properties of different common solvents with respect to asphaltenic compounds are closely related to the surface tension.

We have modeled and predicted the conditions for colloidal effect such as the formation of asphaltene steric colloids formed because of a combination of asphaltenes aggregates / flocs and resins in petroleum fluids through consideration of resins balance / equilibrium between the oil phase and asphaltene-floc-surface phase [33,60,65,66,68].

Colloids are generally unstable and they could deposit. Of course, due to further dilution of a petroleum fluid with light paraffin hydrocarbons, colloids could grow further in size and their deposition will be enhanced (see Figure 18).

Knowing the properties of the fluid and colloids, or suspended solids, we can develop limit-of-Brownian-particle models to predict deposition of colloids from the liquids phase to form a deposited solid phase.

Flocculation and deposition of asphaltene in petroleum fluids can be either reversible or irreversible depending on the nature of asphaltene and resin involved and how far the aggregation and colloid formation and growth has progressed [33,62,68]. Certain experimental observation have indicated a hysteresis when a petroleum fluid conditions are returned to pre-flocculation point of its asphaltene contents. Due to their large size and their adsorption affinity to solid surfaces, asphaltene flocs (random aggregates) can cause quite stable deposits, which may not wash away by current remediation techniques. One of the effective and routine methods of remediation of fresh heavy organic deposits is the use of strong light aromatic solvents, such as xylene, that could dissolve such asphaltene deposits [69]. However, aging of such deposits may prevent it from complete dissolution [54,70].

In order to relate the observed phase transitions in petroleum fluids to the known theory of phase transitions we present the following section.

### 3.3. Theory of Phase-Transitions

From the thermodynamic point of view, phase transitions happen when the free energy of a system is non-analytic for some choice of thermodynamic variables. The distinguishing characteristic of a phase transition is an abrupt change in one or more physical properties, with a small or no change in the intensive thermodynamic variables such as the temperature and pressure. Phase transitions are generally categorized in the first order transitions, the second order transitions and the infinite-order phase-transitions [71].

The first-order phase transitions are those that are associated with an appreciable latent heat. In another words, during a first-order phase transition, a pure compound, as an example, exchanges a fixed and rather large amount of energy in the form of heat. During a first-order phase transition the temperature and pressure of a pure compound will remain constant. Because the latent heat cannot be instantaneously transferred between the compound and its environment, the first-order transitions are associated with "mixed-phase regimes" in which some parts of the system have completed the transition and others have not. Those petroleum fluids phase transitions, which fall into the first-order category, include gas separation from petroleum fluids and crystalline wax formation. It must be pointed out that the first-order gas-liquid transitions in petroleum fluids and other mixtures occur at constant temperature over a range of pressures, as discussed elsewhere in this report.

The second-order phase transitions, also known as the continuous phase transitions are not associated with any latent heat exchange with the surroundings. The classical examples of the second order phase transitions are the ferromagnetic transition, superconductor and the superfluid transition. Examples of the second-order phase transitions in petroleum fluids are asphaltene precipitation from a petroleum fluid due to composition change, asphaltene steric colloid formation, asphaltene deposition from a petroleum fluid due to the limit of Brownian particle size limit and asphaltene micellization [72].

A third category of phase transitions is known as the infinite-order phase transitions. They are known to be continuous but breaking no symmetries. The classical example is the Kosterlitz-Thouless transition [73] in the many quantum phase transitions in two-dimensional electron gases. In the case of petroleum fluids phase transition from asphaltene to asphaltene micelles-coacervates belong to this category of phase transitions.

In Figure 12 we report a qualitative PT diagram of a petroleum fluid, which may experience variety of phase transitions. This diagram as a whole may not apply to all the seven petroleum fluids, but it gives a general picture of how various phase transitions may be distinguished from one another.

To specify all the possible phase-transitions in petroleum fluids we present below the general concept of phase-transitions points.

**3.4. Phase-Transition Points**:

Because of the variety of phase-transitions, which may be occurring in petroleum fluids due to variations in temperature, pressure and composition, we observe a number of specific phase-transition points. Specifically there are up to eleven different and well-defined phase-transition points, which could occur in a petroleum fluid.

These eleven points are either observable directly by the naked eye or they can be distinguished by indirect property change indicators. In Table 2 we list these known and well-defined phase-transition points. Obviously not every petroleum fluid will possess all these eleven phase-transition points. The mere fact that a petroleum fluid will exhibit certain of these phase-transition points will distinguish it from the other kinds of petroleum fluids. In what follows we give a brief description of the eleven phase-transition points (PTP's) and the kind of petroleum fluid in which they may appear.

*PTP # 1*. The critical solution point is where the vapor and liquid phases separated from a petroleum fluid possess the same compositions, pressures, temperatures and densities. Generally, all liquid mixtures possess a critical solution point.

*PTP # 2*. Bubble point is the onset point of phase-transition from a liquid to a vapor phase upon increase of the temperature and/or lowering the pressure of the liquid phase.

*PTP # 3*. Dew point is the onset point of phase-transition from a vapor to a liquid phase upon decrease of the temperature and/or increasing the pressure of the vapor phase.

*PTP # 4*. Cloud point is the onset point of crystalline solid formation in petroleum fluids upon lowering its temperature. This phase-transition point is an indicator of the wax content of the petroleum fluid.

**Table 2** – *Known and well-defined phase-transition points in petroleum fluids*

| PTP # | Phase-Transition Point | *Phases involved* | Components involved | Major variables | Indicator | Ref. # |
|---|---|---|---|---|---|---|
| 1 | Critical solution point | Vapor-Liquid | Hydrocarbons | Pressure, Temperature | Visual | [10] |
| 2 | Dew point | Vapor-Liquid | Hydrocarbons | Pressure, Temperature | Visual | [10] |
| 3 | Bubble point | Vapor-Liquid | Hydrocarbons | Pressure, Temperature | Visual | [10] |
| 4 | Cloud point | Liquid-Solid | Paraffins / Wax | Temperature | Viscosity, etc. | [34] |
| 5 | Static Pour point | Liquid-Solid | Paraffins / Wax | Temperature, viscosity | Viscosity, Velocity | [34] |
| 6 | Dynamic Pour point | Liquid-Solid | Paraffins / Wax | Temperature | Visual, Viscosity | [34,74] |
| 7 | Onset of asphaltene precipitation | Liquid-Solid | Asphaltene | Paraffins, Pressure | Oil-Water IFT | [57] |
| 8 | Onset of asphaltene steric colloid formation | Liquid-Liquid | Asphaltene, Resin | Paraffins, Pressure | Viscosity | [75] |
| 9 | Onset of asphaltene deposition | Liquid-Solid | Asphaltene, Resin | Paraffins, Pressure | Visual, Microscopy | [74] |
| 10 | Onset of asphaltene micellization | Liquid- Liquid | Asphaltene | Aromatics | Viscosity, Surface tension | [53,55] |
| 11 | Onset of asphaltene micelle coacervation | Liquid-Liquid | Asphaltene micelles | Aromatics | Viscosity, Surface tension | [53,55] |

*PTP # 5*. The static pour point is the lowest temperature at which a petroleum fluid will cease to flow out of a tilted test jar. Pour point is predominantly known to be due to the crystal growth and connectivity of wax crystals, which then increases the viscosity of the petroleum fluid extremely high preventing the flow of the oil.

*PTP # 6*. Dynamic pour point is the lowest temperature at which a petroleum fluid will cease to flow out of a tilted, but continuously agitated, test jar. It occurs at a lower temperature than the static pour point. The difference of the static and dynamic pour point temperatures is a measure of the nature of the paraffin wax present in a petroleum fluid. There is not much data available yet for petroleum fluids dynamics pour point and its measurement procedure is not yet standardized.

Petroleum fluids containing asphaltenes and resins could have five additional phase-transition points depending on their percentages of paraffins, aromatics, asphaltenes, resins as well as the nature and molecular weights of these compounds. These five additional phase-transition points, as described in Table 2 are:

*PTP # 7*. Onset of asphaltene precipitation

*PTP #* **8**. Onset of asphaltene steric colloid formation
*PTP #* **9**. Onset of asphaltene deposition
*PTP #* **10**. Onset of asphaltene micellization
*PTP #* **11**. Onset of asphaltene micelle coacervation

We have discussed the details of the latter five phase-transitions points in our earlier publications [72]. While these five phase-transition points may not be visually observable, they define the behavior of petroleum fluids in regard to the solid phase formation and separation.

Except for the critical point, bubble point line and dew point line, which belong to light hydrocarbon components of petroleum fluids, the other eight phase-transition points are due to the heavy components. The latter eight phase-transition points represent the onset of heavy organics (wax, asphaltene, resin, etc.) transformations and separations from petroleum fluids. The knowledge about all these phase-transition points will allow us to understand and quantify the nature of petroleum fluids and develop models to predict and strategies to control various phase separations including the deposition of heavy organics from petroleum fluids. In Figure 12 we report the locations of phase-transition points 1-9 as discussed above and reported in Table 2.

## 4. Discussion

There is a great deal of interest and need in the petroleum and natural gas industries to develop computational packages to predict the phase behaviors of petroleum fluids. This is because the knowledge about petroleum fluids phase behavior will allow us to design the production, transportation and utilization systems efficiently and will help us to prevent problems associated with fouling in petroleum fluids arteries due to untimely heavy organics deposition.

Our knowledge about phase behavior occurring in natural gas, gas-condensate and light crude is quite complete thanks to the extensive amount of database and computational packages which have been developed in the past century. However, worldwide shortage of such light petroleum fluids has caused the need to use not only intermediate crudes, which are also rather scares, but heavy oil, tar sand and oil shale. For efficient utilization of such heavy petroleum fluids we need to understand and quantify their phase behavior.

In this report we have identified all the phase transitions, which could occur in the seven naturally occurring hydrocarbon fluids known as petroleum fluids. We have also specified eleven phase transition points, which are responsible for all such phase transitions. Also in this report the nature of every petroleum fluid is presented and their constituents including their non-hydrocarbon components, known as impurities, are identified and categorized. It is shown that the heavy fractions in petroleum fluids are generally made of continuous families of constituents. Their major components, which have a strong role in their phase behavior, include wax, diamondoids, asphaltenes, heavy aromatics and petroleum resins.

We have introduced the generalized petroleum fluids phase behavior in light of the known theory of first-order, second-order and infinite-order phase transitions. We have also discussed about the effects of variations of composition, temperature and pressure on the phase behavior of petroleum fluids. We have suggested eleven distinct phase-transition points of petroleum fluids, which are related to various state variables, and constituents of petroleum fluids. The aim of this report is to generalize and relate phase behaviors of all the seven naturally occurring petroleum fluids into a unified perspective. The major requirement in the efficient design of production, transportation and processes dealing with petroleum fluids is the accurate and efficient prediction of their phase behavior and other properties as they change phases between vapor, liquid and solid.

The concepts introduced in this report will be necessary for development of a comprehensive package to predict various seven petroleum fluids behavior as well as the behavior of their blends, which is of major interest in the petroleum industry. The challenges ahead in terms of generating such unified modelling are the following:

    i. To relate all the eleven phase transition points to microscopic phenomena and come up with predictive methods for each.

    ii. To produce a unified theory of phase transitions which will take into account changes of phases of molecules and their assembly between, not only liquids and vapour phases, but also solid crystalline, solid aggregates, colloid and micelle formations and micelle and colloid coacervation or self assembly.

    iii. To develop a multi-scale model of phase behavior predictions for a multi-component / multi-family polydisperse mixture which will have its input certain compositions and/or sets of data from the eleven phase transition points.

    iv. Equations of state are useful to predict phase equilibrium between fluid phases. However, for prediction of solid phase formation, aggregations (micelles, colloids and coacervates) other theories will be necessary. A combination of such phase transitions may be of reversible or irreversible nature as discussed above.

It is with no doubt that the techniques and facilities being developed for nanotechnology [76] can be quite useful to provide us a better understanding of the molecular interactions and self-assemblies, which are occurring during in petroleum fluids during various phase transitions.

**Glossary**

*aggregation:* Attachment due to attraction between particles (polymers, macromolecules, etc.) due to non-covalent or covalent forces.

*atomic force microscope:* a microscope with extreme high resolution measuring intermolecular forces and producing nano-scale images.

*bubble point:* The onset point of phase-transition from a liquid to a vapor phase upon increase of the temperature and/or lowering the pressure of the liquid phase.

*cloud point:* the onset point of crystalline solid formation in petroleum fluids upon lowering its temperature. This phase-transition point is an indicator of the wax content of the petroleum fluid.

*coacervate:* self-assembly of micelles

*colloid:* a collection of particles of a substance aggregated and suspended in a fluid media.

*gas condensate:* a mixture of low to intermediate molecular weight hydrocarbons which is in liquid state at normal conditions and is in gas phase at high pressures due to dissolution in supercritical natural gas.

*critical solution point:* Where the vapor and liquid phases separated from a petroleum fluid possess the same compositions, pressures, temperatures and densities. Generally, all liquid mixtures possess a critical solution point.

*dew point:* the onset point of phase-transition from a vapor to a liquid phase upon decrease of the temperature and/or increasing the pressure of the vapor phase.

*diamondoids:* cage-shape saturated hydrocarbons as presented in Section 2.3.2 of this chapter.

*dynamic pour point:* the lowest temperature at which a petroleum fluid will cease to flow out of a tilted, but continuously agitated, test jar. It occurs at a lower temperature than the static pour point. The difference of the static and dynamic pour point temperatures is a measure of the nature of the paraffin wax present in a petroleum fluid. There is not much data available yet for petroleum fluids dynamics pour point and its measurement procedure is not yet standardized.

*first order phase transition:* a phase transition during which appreciable latent heat is exchanged between the substance and its surroundings.

*flocculation:* The same as aggregation.

*FRACTAL:* an object, shape or quantity is considered a FRACTAL if it exhibits self-similarity in various sizes and/or quantities. A FRACTAL entity is said to have a dimension which is not necessarily an integer as are Euclidean dimensions.

*infinite-order phase transition:* continuous phase transitions which break no symmetries.

*micelle:* self-assembly of surfactant molecules dispersed in a liquid medium.

*phase transition points:* a point on the phase diagram bordering the two phases.

*polydisperse fluid:* a fluid containing various families of molecules, each with a size distribution.

*retrograde condensation:* separation of a liquid phase from a high pressure gaseous mixture upon expansion.

*second order phase transition:* a phase transition during which little or no latent heat is exchanged between the substance and its surroundings.

*static pour point:* the lowest temperature at which a petroleum fluid will cease to flow out of a tilted test jar. Pour point is predominantly known to be due to the crystal growth and connectivity of wax crystals, which then increases the viscosity of the petroleum fluid extremely high preventing the flow of the oil.

*Steric colloid:* an aggregate of a macromolecular substance (asphaltene in the present case) peptized and suspended in a fluid media by a polar but smaller molecule (petroleum resins in the present case).

**Nomenclature**

## Biographical Sketch of the author

**G.Ali Mansoori** received BSc-*ChE* degree from University of Tehran, MSc-*ChE* degree from University of Minnesota, PhD from the University of Oklahoma and his postdoctoral training at Rice University, the latter in 1970. He is a professor of bioengineering, chemical engineering and physics at the University of Illinois at Chicago. He has been a visiting professor at University of Pisa, ITB, University of Kashan, Sharif University of Technology; and a visiting scientist at the Argonne National Laboratory, National Institute of Standards and Technology and CNR-Pisa. He has been a consultant to ARCO, BJ Services, British Petroleum (BP), C3 Int'l-LLC, Chevron Oil Field Services, DuPont, Exxon, Eng. Research Corp., Federation of American Scientists, Harza Eng. Co., Hitachi, Ltd., Hydrotech, IMP, Institute of Catalysis of Novosibirsk-Russia, Mitsubishi Chemicals, Motorola, MUCIA, NIGC, NIOC, Norsk Hydro, PEMEX, PETROBRAS, PetroStat Laboratories, Science Applications., Shell Int'l B.V., Synethic/Johnson-Matthey, Technology International., The Anián Community (Reuters), United Nations

(UNDP; TOKTEN Project), UOP-LLC, Vista Research. Dr. Mansoori's academic and professional honors include: Academician of the International Academy of Creative Endeavors (science, arts, social issues); Algorithm scientific international award; diploma of honor from Pi Epsilon Tau National Petroleum Engineering Honor Society; honorary academician of the International Eco Energy Academy; Kapitsa gold medal from the Russian Academy of Natural Sciences; medal of fundamental science from UNESCO; recognition of dedicated service award from the fuels & petrochemicals division of AIChE, science medal from *Vezarat a Farhang va Aamozesh Aali*, honorary member of the IRI National Academy of Sciences, undergraduate instructional award from UIC. His research and educational activities include arterial blockage / fouling in petroleum and natural gas industries, atomic and molecular nanotechnology, molecular based study of condensed matter, disease diagnostic methods and therapeutic agents, nanostructures design (nanoclusters, nanoconjugates, nanoparticles), phase transitions, *ab initio* methods, density functional and molecular dynamics simulations, statistical mechanics, thermodynamics. Prof. Mansoori has published over 350 technical papers including seven books.